# Turning the tables in citation analysis one more time:

# Principles for comparing sets of documents

*Journal of the American Society for Information Science and Technology* (in press)

Loet Leydesdorff,[a] Lutz Bornmann,[b] Rüdiger Mutz,[c] and Tobias Opthof [d,e]

[a] Amsterdam School of Communication Research (ASCoR), University of Amsterdam, Kloveniersburgwal 48, NL-1012 CX Amsterdam, The Netherlands; loet@leydesdorff.net

[b] Max Planck Society, Hofgartenstrasse 8, D-80539 Munich, Germany; bornmann@gv.mpg.de

[c] ETH Zurich, Professorship for Social Psychology and Research on Higher Education, Zähringerstrasse 24, CH-8092 Zurich, Switzerland;

[d] Experimental Cardiology Group, Center for Heart Failure Research, Academic Medical Center K2-105, 1105 AZ Amsterdam, The Netherlands; t.opthof@inter.nl.net

[e] Department of Medical Physiology, University Medical Center Utrecht, 3508 CM Utrecht, The Netherlands.

abstract
**Abstract**

We submit newly developed citation impact indicators based not on arithmetic averages of citations but on percentile ranks. Citation distributions are—as a rule—highly skewed and should not be arithmetically averaged. With percentile ranks, the citation of each paper is rated in terms of its percentile in the citation distribution. The percentile ranks approach allows for the formulation of a more abstract indicator scheme that can be used to organize and/or schematize different impact indicators according to three degrees of freedom: the selection of the reference sets, the evaluation criteria, and the choice of whether or not to define the publication sets as independent. Bibliometric data of seven principal investigators (PIs) of the Academic Medical Center of the University of Amsterdam is used as an exemplary data set. We demonstrate that the proposed indicators [$R(6)$, $R(100)$, $R(6,k)$, $R(100,k)$] are an improvement of averages-based indicators because one can account for the shape of the distributions of citations over papers.

**Introduction**

In a critique of the normalization of citation indicators in use with well-established indicators such as the so-called "crown-indicator" (*CPP/FCSm*) of the Leiden Center for Science and Technology Studies (CWTS),[1] the MNCR of the Flemish evaluation center ECOOM in Louvain (Glänzel *et al.,* 2009, at p. 182), and the MOCR/MECR measure used by the ISSRU in Budapest (Schubert & Braun, 1993 and 1996),[2] Opthof & Leydesdorff (2010) repeated a point that had been made previously by Lundberg (2007), namely that these indicators first aggregate the numerator and the denominator separately, and then normalize by dividing the two means as a ratio. However, this can be considered as a transgression of the order of operations: one should first normalize for each individual paper against a reference set and then average over the resulting distribution of ratios (or use the median or any other statistics of this distribution). This sequence produces a mathematically consistent indicator (Gingras & Larivière, 2011; Waltman *et al.*, 2011) with standard error terms that enable one to determine whether differences between document sets are statistically significant.

The controversy (Bornmann, 2010; Leydesdorff & Opthof, 2010 and 2011; Van Raan *et al.*, 2010a; Waltman *et al.*, 2011) led CWTS to propose a new crown indicator "*MNCS*" (the Mean Normalized Citation Score) and several derivatives of this indicator (such as *MNCS1* and *MNCS2*; Van Raan *et al.*, 2010b; Waltman *et al.*, in press). Using *MNCS*, the "rate of averages" (*CPP/FCSm*) is replaced with the "averaging of rates"—as Gingras & Larivière (2011) summarized the core issue of this debate. The "new crown indicator" does not suffer from the shortcomings of the old one (Waltman *et al*., 2011). In our opinion, however, several issues which can be raised with respect to the normalization of citation scores have not yet received sufficient attention. Closure of the debate by means of establishing a new "crown indicator" might from this perspective be premature.

---

[1] *CPP/FCSm* is defined as the *mean* citation score of the document set under study divided by the *mean* of the citation score of the journals representing this field according to the ISI Subject Categories of Thomson Reuters.
[2] *MNCR* means "mean normalized citation rate"; *MOCR/MECR*: the "mean of the observed citation rate divided by the mean of the expected citation rate."

First, the new crown indicator (like the old one) is based on using (arithmetic) averages of—as a rule—highly skewed citation distributions (e.g., Albarrán & Ruiz-Castillo, 2011; Boyack, 2004, p. 5194; Seglen, 1992). Both Bornmann & Mutz (2011) and Leydesdorff & Opthof (2011) raised the issue that it might be better to use medians and non-parametric statistics instead. More specifically, Bornmann & Mutz (2011) proposed an elaboration into a scheme which they called the "percentile rank approach." This non-parametric approach is already in use as the evaluation scheme in the *Science & Engineering Indicators* of the National Science Foundation of the USA (National Science Board, 2010), prepared biannually by the American corporation ipIQ. In this scheme the focus is not only on (relative) citation rates, but also on the top-cited papers (Bornmann *et al*., 2010a).

Bornmann *et al*. (2008) raised the issue of using journals or groups of journals (aggregated into so-called Subject Categories by the Institute of Scientific Information (ISI) of Thomson Reuters) as systems of reference for the normalization. Rafols & Leydesdorff (2009) argued that these ISI Subject Categories were developed for reasons other than bibliometric measurement and had been estimated faulty in more than 40% of individual attributions (Boyack *et al*., 2005; Garfield & Pudovkin, 2002, at p. 1113n.). The use of these categories for journal classification, as in the field-normalization of many of the existing indicators (including *MNCS*), might therefore be unfortunate. The ECOOM center in Leuven (Belgium) developed its own classification scheme (Glänzel & Schubert, 2003), but Rafols & Leydesdorff (2009) showed that this classification of journals does not improve on the ISI Subject Categories; the latter are finer grained and therefore less error-prone than the newly proposed ones (Leydesdorff & Rafols, 2009).

In addition to classifications of journals grouping potentially heterogeneous sets, journals themselves can be heterogeneous in terms of document types, citation half-lives, cognitive substance, etc. (Leydesdorff, 2008; Moed, 2010). Bornmann *et al.* (2008) proposed using classification schemes at the level of individual papers such as the Medical Subject Headings (MeSH) of Medline, the publicly available database of the National Institute of Health of the USA. Bornmann *et al.* (2011; in press) applied the percentile rank approach using the classifications of *Chemical Abstracts*. Leydesdorff & Opthof (2010) proposed appreciating differences among individual papers by using fractional counting of citations in terms of the number of references in the *citing* papers, arguing that differences in so-called "citation potentials" (Garfield, 1979, at p. 365) are generated on this side of the citation process. Similar proposals have been made by Moed (2010), Zitt (2010), and Zitt & Small (2008) for the normalization of journal impacts using citing-side normalizations (Leydesdorff & Bornmann, 2011).

In this study, we focus on *cited-side* normalizations and try to take the discussion one step further by raising, in addition to the problem of normalization, the problem of evaluation (for example, in terms of the 1% most highly cited papers). Furthermore, we address the issue of how to normalize for differences in productivity (publication rates) when comparing citation distributions among document sets. With the notable exception of the *h*-index—which is also defined in terms of numbers of publications that meet a specified criterion (Hirsch, 2005)—citation indicators have hitherto not paid sufficient attention to the effects of productivity rates on the evaluation indicators.

Lundberg (2007, at p. 148) noted that one does not have to average the (field-normalized) citation scores, but can also use their sum values as a "total" field-normalized citation score. Using the Leiden Rankings, CWTS multiplies the product of the number of publications *P* with the old crown indicator *CPP/FCSm* in order to obtain as a result the so-called "brute force indicator". In the new set, analogously, a "total normalized citation score" was distinguished (Van Raan et al., 2010b, at p. 291), but hitherto this indicator was not yet used for the evaluation. None of these indicators enable us to answer questions such as how to weigh one paper in the top-1% range against five (or more) papers in the top-5% range, etc. The current schemes do not allow for quantitative assessment of such comparisons. Yet, these questions are most pressuring when one wishes to use bibliometric evaluations for distinguishing between "good" and "excellent" research for reasons of policy-making or institutional management (Opthof & Leydesdorff, in preparation).

In summary, we distinguish first a number of analytical questions and then elaborate on the percentile rank approach to develop a set of criteria to be met by this new indicator for citation analysis. Let us list these criteria:

1. A citation-based indicator must be defined so that the choice of the reference set(s) (e.g, journals, fields) can be varied by the analyst independently of the question of the

2. The citation indicator should accommodate various evaluation schemes, for example, by funding agencies. Some agencies may be interested in the top-1% (e.g., National Science Board, 2010) while others may be interested in whether papers based on research funded by a given agency perform significantly better than comparable non-funded ones (e.g., Bornmann *et al.*, 2010b);
3. The indicator should allow productivity to be taken into account. One should, for example, be able to compare two papers in the $39^{th}$ percentile with a single paper in the $78^{th}$ percentile (with or without weighting the differences in rank in an evaluation scheme as specified under 2.);
4. The indicator should provide the user, among other things, with a relatively straightforward criterion for the ranking (for example, a percentage of a maximum) that can then be tested for its statistical significance in relation to comparable (sets of) papers;
5. It should be possible to calculate the statistical error of the measurement.

In this study, to progress to these stated objectives, our two teams, previously involved independently in the controversy, have joined forces. First, we replicated the measurements of CWTS (2008) and Opthof & Leydesdorff (2010) for the purpose of establishing the percentile ranks of citations of the papers under study in their respective reference sets, and secondly, we elaborate on the ideas of Bornmann & Mutz (2011) to develop percentile ranks as schemes which enable us to compare across sets using non-parametric statistics. Using the percentile rank values allows us to express differences in terms of numbers which can be considered as percentages, and we will specify how differences among these numbers can be tested for their significance. Using the six-rank scheme of the National Science Foundation (Bornmann *et al.*, 2010a; National Science Board, 2010) as an example, we show the effect of the non-linear transformation implied when using such an evaluation scheme.

**Methods and materials**

Because, as academics, we do not have the means to manipulate yearly volumes of the *Science Citation Index* like the centers that license the database for evaluation purposes, we have used the Web-of-Science interface at the Internet and confined the normalization to seven sets and the comparable documents (in terms of document types) in the same journals and publication years. Although the choice of using journals or fields for the normalization matters (Colliander & Ahlgren, 2011), the specific normalization is not fundamental to our analytical argument, but serves us as an example. Our scheme requires one normalization or another against a reference set for each paper (Radicchi *et al.*, 2008). (One could, for example, consider the un-normalized citation rates as a zero-normalization because all reference sets are then set equal to unity.)

Because the ISI split the category of "articles" into articles and proceedings papers in the period under study (in October 2008), we will consider "articles OR proceedings papers" as our reference sets in the publishing journals in the specific years of publication of 241 source documents. These source documents were published by seven principal investigators (PIs) in the Academic Medical Center (AMC) of the University of Amsterdam. The PIs belong to a group of 232 scientists evaluated by CWTS (2008 and 2010). Opthof & Leydesdorff (2010) provide reasons for selecting these seven scientists in terms of the distributions of citations as a representative sample given the performance range in the larger group. The seven authors published 23, 37, 22, 32, 37, 65, and 32 papers, respectively, during this period. The seven document sets overlap in seven coauthored papers. Thus, 248 – 7 = 241 papers could be attributed to seven document sets. The seven sets constitute our units of evaluation.

For these 241 documents and their corresponding reference sets in the journals published in the same years, we determined citation rates in early November 2010. For each paper thus a number of citations per paper ("CPP" in the terminology of CWTS) and "journal citation score" ("JCS") can be computed, and for each set accordingly a so-called *CPP/JCSm* (mean citation rate per paper divided by the mean journal citation score) can be calculated both in terms of a "rate of averages" or as an "average of rates" (that is, an MNCS-type of indicator but then defined at the journal level; Van Raan *et al*., 2010b, at p. 291).

In order to move to the percentile rank approach, the citation of each paper is rated in terms of its percentile in the distribution. In each reference set, the number of papers with citations fewer than (<) the citation of a paper *i* is expressed as a percentage. (Tied, that is, precisely equal, numbers of citations thus are not counted as "fewer than.") The percentiles are then rounded as integers. In other words: if 65.4 % of the papers were below that of the $i^{th}$ paper with a certain citation, then the percentile score of this paper would be classified into the $65^{th}$ percentile class.[3] Thus, for each set under study a column vector with 100 values (from the $0^{th}$ to the $99^{th}$ percentile, but with ranks 1 to 100) is created. Note that the seven column vectors—representing the seven sets—are now equal in size and thus comparable.

---

[3] If a journal publishes only a single review each month, the maximum percentile score using this scheme would be in the $91^{st}$ percentile rank since 11/12 = 91.7. A different approach to determining percentile ranks is provided by Pudovkin & Garfield (2009).

From this matrix (7 columns each with 100 rows), the six percentile impact classes used by the NSF (National Science Board, 2010; cf. Bornmann *et al*., 2010a) for the evaluation were aggregated as follows:

(1) bottom-50% (papers with a percentile less than the 50$^{th}$ percentile),

(2) 50$^{th}$ – 75$^{th}$ (papers within the [50$^{th}$; 75$^{th}$[ percentile interval),

(3) 75$^{th}$ – 90$^{th}$ (papers within the [75$^{th}$; 90$^{th}$[ percentile interval),

(4) 90$^{th}$ – 95$^{th}$ (papers within the [90$^{th}$; 95$^{th}$[ percentile interval),

(5) 95$^{th}$ – 99$^{th}$ (within the [95$^{th}$; 99$^{th}$[ percentile interval),

(6) top-1% (papers with a percentile equal to or greater than the 99$^{th}$ percentile).

Thus, a contingency table of seven scientists (independent variable) and six categories (dependent variable) is generated. Note that the scores in this matrix are non-parametric and ordinal-scaled while the previous ones of hundred percentile classes are also non-parametric, but interval-scaled. In other words, the transformation by aggregation into six classes is non-linear: the percentile scores are transformed for the purpose of a normative evaluation. We use the evaluative scheme of the NSF in this study as an example of such an evaluation scheme.

The mean percentile rank scores are calculated by weighting the relative frequencies (i.e., probabilities) $p(x)$ in each set $k$ with their rank $x$, as follows:

$$R(i) = \sum_{x=1}^{i} x \cdot p(x) \tag{1}$$

In the case of six ranks $i = 6$, and in the case of hundred $i = 100$, etc. Note that one paper in the 78th percentile weighs twice as much as a paper in the 39th percentile in the case of the hundred percentile ranks, while in the case of six ranks the paper in the 78th percentile would count three times as much as a paper in the 39th percentile—or equally as three papers in the 13th percentile. The maximum weight in the case of 100 classes [R(100)] is 100, while this maximum is six in the case of six classes [$R(6)$]. The minimum is always 1, that is, when all papers are to be placed in the first (and lowest) category.[4]

Citation performance above or below a medium level can be evaluated by testing, for example, the 100 percentile ranks against a median value of 50 using Wilcoxon's signed-rank test (which is available under the non-parametric tests in SPSS). The citation performance of a single scientist can further be tested against a reference value for $R(6)$. This latter reference value can be obtained by the sum of the products of proportions of the percentile classes (50:25:15:5:4:1) multiplied with the rank numbers of the classes: for example, each count in the bottom-50% class counts as one, and a count in the top-1% as six. One thus obtains an expected value of $R(6)$ for the case of random attribution, as follows: 0.50*1 + 0.25*2 + 0.15*3 + 0.05*4 + 0.04*5 + 0.01*6 = 1.91. The performance of a scientist with an $R(6)$ at about 1.91 is at the medium-level. There is a maximal possible citation performance with an $R(6)$ of 6 (all papers belong to the top-1%: 6 x 1) and a minimal possible citation performance with an $R(6)$ of 1 (all papers belong to the bottom-50%: 1 x 1). The observed distribution of the papers of a single scientist over the six percentile rank classes can be tested against the expected distribution (50:25:15:5:4:1) using $\chi^2$-statistics.

---
[4] One may wish to leave the papers without citations (hitherto) out of the analysis, but they were included in the 0th percentile in this study.

In the case of both $R(6)$ or $R(100)$—or any other scheme for the evaluation—the seven document sets can be tested against each other for statistical significance of the differences using Dunn's test or Mann-Whitney's U test. Confidence levels can also be indicated.[5] First, one should test whether differences among the scientists under study are significant using Kruskal-Wallis (rank variance analysis). If the null hypothesis is not rejected (that is, no significant differences among the sets are found), then the analysis should be ended here because it is not relevant to test further. In the other case, one can test the differences using the Mann-Whitney U test on each two samples, or Dunn's test including an *ex-post* Bonferroni correction for multiple comparisons. The Mann-Whitney U test is more conservative—that is, less inclined to flag differences as significant—than Dunn's test, the non-parametric version of the ANOVA-based post hoc test (Levine, 1991, pp. 68 ff.). Since Opthof & Leydesdorff (2010) used the latter test, we have stayed with this choice.

Due to the facts that (1) the citation data considered are not normally distributed, (2) the variances of the data are not homogeneous across scientists, and (3) the dependent variables, especially the percentile impact classes, are not continuous but ordinal, we opt for non-parametric instead of parametric statistical procedures (Corder & Foreman, 2009; Kvam & Vidakovic, 2007; Sheskin, 2007). The non-parametric statistics do not make strong assumptions with respect to the distribution of the data. For consistency reasons, we use non-parametric statistics throughout this study. One can use Dunn's test for the *ex-post* correction (e.g., using ANOVA in SPSS), but one has to correct for the overall so-called family-wise alpha-error accumulation (Type I) across all possible pairwise comparisons (Levine, 1991, pp. 68 ff.). This error increases with the number of these pairwise comparisons *c*. For each pairwise comparison an adjusted alpha error of 0.05 divided by the number of all possible pairwise comparisons was used instead of an alpha error of 0.05. For n=7 there are $c=n*(n-1)/2=7*6/2=21$ comparisons, and the adjusted alpha-level therefore amounts to 0.05/21=0.0023. In general, this alpha-level of 0.05/c can be used as the significance level with the *ex-post* correction. Additionally, one has to assume no dependency among observations due, for example, to multiple publications on the same topic.

---

[5] SPSS (v. 18) provides confidence levels when comparing means, but also non-parametrically when using Dunn's test with adjusted alpha-levels (see below). Alternatively, one can compute a Goldstein-adjusted confidence interval as equal to 1.396 times the standard error of the mean. Non-overlapping differences between each two sets can then be estimated as significance of this difference at the 5% level (Goldstein & Healy, 1995).

Another issue is the normalization of the relative frequencies *p*(x) in terms of the respective margin totals ($p_i = f_i / n_i$; $n_i = \Sigma_i f_i$). If a scientist with 10 publications had one publication in the 99$^{th}$ percentile, this publication adds 1/10$^{th}$ times 100—the rank number—or 10 percentage points to his/her citation rank profile *R*(100) given the probability mass function in Equation 1 (above). However, a scientist with 100 publications but only one in the 99$^{th}$ percentile would add only a single percentage point to this score. (An analogous reasoning can be elaborated for *R*(6).) Since publications in the highest ranks are scarce—given the well-known skewness in empirical citation distributions (e.g., Albarrán & Ruiz-Castillo, 2011; Seglen, 1992)—this system can thus be expected to disadvantage productive scientists.

This effect disappears when the frequencies are not calculated relative to each subset (e.g., the *œuvre* of each scientist), but to the total set under study ($N = \sum_{k=1}^{k} n_k$ ; in our case $k = 7$ document sets). The weighting is then similar for each scientist in the aggregated set. In order to make the resulting ranks comparable with those individually weighted (for example, as percentages), the results have to be multiplied again with $k$ (in our case, $k = 7$). We distinguish between the two normalizations by writing below $R(6)$ and $R(100)$ when normalizing over the six categories or 100 percentile classes for each document set (vector) as independent samples, and $R(6,k)$ and $R(100,k)$ when normalizing also over the second dimension of the $k$ subsets of a single sample.

Using $R(100,k)$, the uncontrolled effects of differences in publication rates on the means and the medians of the impact indicators (cf. Rousseau & Leydesdorff, 2011) are completely taken out of the equation: each publication in our case of the 248 documents under study has a weight of (1/248) in its percentile rank. The resulting percentile ranks [$R(100,k)$] can thus be compared directly with one another across the sets: two publications in the 39<sup>th</sup> percentile in one set now weigh as much as one publication in the 78<sup>th</sup> percentile in *another* using $R(100,k)$. Using $R(6,k)$ a non-linear transformation is involved, but papers in the same rank are equally appreciated using both $R(6,k)$ and $R(100,k)$. We will discuss the differences of and similarities between the two normalizations (in terms of relative frequencies) in the next section in empirical terms.

In summary, we have thus constructed an indicator in which the different criteria specified above can be distinguished analytically. The weighing is a technical consequence of the normative decision for one evaluation scheme or another; for example, in terms of six or hundred percentile rank classes. This normalization specifies aggregation rules at the level of sets. The normalization in terms of reference sets is determined at the level of individual papers; for example, against the set of all papers in the same journal in the same year and of the same document type. This latter normalization implies a decision on substantive grounds provided, for example, in the discourse or by the state of the art in bibliometrics (Leydesdorff & Opthof, 2011).

Each set can be attributed a comparable score between 1 and 100 in the case of $R(100)$ and $R(100,k)$ or between 1 and 6 in the case of $R(6)$ and $R(6,k)$ (The latter score can also be expressed as a percentage of six). The six classes or any other normative scheme for the assessment can be derived directly from the matrix of the hundred percentile values because the evaluative scale is based on specific aggregation rules which can be chosen differently depending on the purposes of the evaluation.

One disadvantage of our scores might seem to be that the idea of a "world average" provided by the old "crown indicator" as a baseline ($CPP/FCSm = 1$) has to be abandoned. In our opinion, an average is always sample-dependent unless one knows the population. The sample of documents can be as large as all documents contained in the *Science Citation Index*, but also the latter remains a sample which is based, for example, on Garfield's (1971) Law of Concentration. The "new crown indicator" of CWTS, for example, defines the world average as the average citation scores in the corresponding ISI Subject Categories of the journals in which the publications are published after controlling for document types and publication years (cf. Leydesdorff & Opthof, 2011).

More importantly, the concept of a "world average" as an evaluation standard confounds the three analytically independent degrees of freedom of (1) external normalization against a reference set for each paper, (2) normalization within sets and subsets in order to be able to apply statistics, and (3) evaluation standards. By the seemingly attractive integration into a single number, one risks loosing the possibility of using statistics and therefore the indication of error. Instead, CWTS and ECOOM used "rules of thumb" to indicate significance in the deviation from the world standard as 0.5 (Van Raan, 2005) or 0.2 (CWTS, 2008, at p. 7; cf. Schubert & Glänzel, 1983; Glänzel, 1992 and 2010).[6]

The statistics implied in our procedures may seem sophisticated and at first sight complex because they involve non-parametric routines. When fully elaborated, these statistics can be automated in SPSS as a batch job. In this study, however, we guide the reader step-by-step through the possible procedures using relatively small sets as an example in order to enable users to reproduce the percentile rank evaluation using their own datasets, their reference sets, and potentially different evaluation schemes.

**Results**

The distributions of the 100%-percentiles for the papers of the seven scientists under study are shown in Figure 1, both as scatter plots and box plots. The black dots in the boxes represent the arithmetic mean, the stars the minima and maxima, respectively. The borders of the boxes indicate the 25%, 50%, and 75% quartiles of individual distributions. Obviously, all scientists score across the whole variance, but in some cases (e.g., Scientists 2 and 3) the concentration in the top half is larger than at the bottom. Scientists 5 and 6 have publications in the 0[th] percentile. For Scientist 5 these concerned papers that had not been cited; for Scientist 6 poorly cited ones.

---

[6] Schubert & Glänzel (1983) based their reasoning on normal distributions (Glänzel, 2010). The reasoning can be used to estimate error in large sets (Glänzel, *personal communication*, 16 November 2009), but this estimator lacks sufficient precision for evaluations of smaller sets.

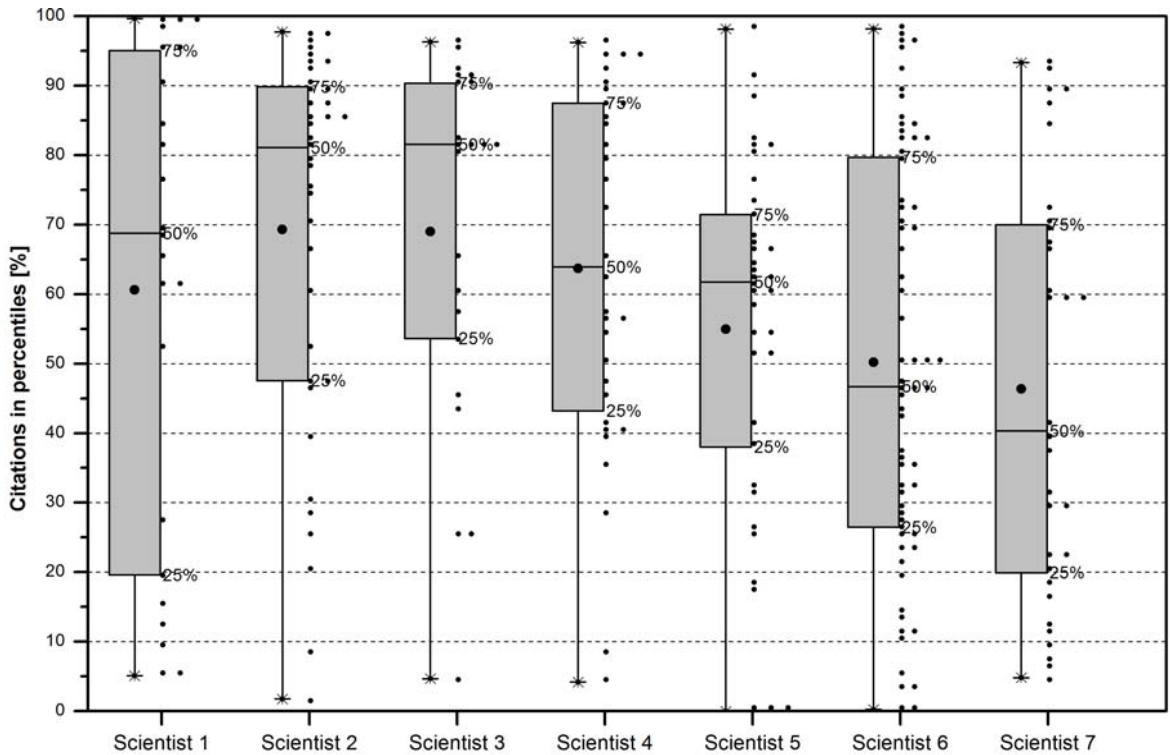

**Figure 1**. Boxplots for citations of each paper in percentiles separated for seven scientists. The black dot in the box represents the arithmetic mean, the stars the minimum and maximum, respectively. The borders of the boxes indicate the 25%, 50%, and 75% quartiles of the individual distribution.

The ordering of the scientists from one to seven was based by Opthof & Leydesdorff (2010) on the ranking of these scientists in the original report of CWTS (2008). These rankings were based on *CPP/JCSm* in the CWTS terminology which can be defined as:

$$Avg(CPP)/Avg(JCS) = \frac{\sum_{i=1}^{n} c_i / n}{\sum_{i=1}^{n} e_i / n} = \frac{\sum_{i=1}^{n} c_i}{\sum_{i=1}^{n} e_i}$$

In this formula, $c_i$ denotes the number of citations of a document $i$, $e_i$ the number of expected citations on the basis of the reference set, and $n$ the number of documents. In this case, each reference set is the set of publications in the same journal with the same publication year and of the same document type. Table 1 provides first the replication of this *CPP/JCSm* on the basis of our downloads of data (in November 2010)[7] and then in the third column the ranking based on the alternative indicator proposed by Opthof & Leydesdorff (2010; cf. Gingras & Larivière, 2011; Lundberg, 2007; Van Raan *et al*., 2010, at p. 291; Waltmann *et al*., 2011):

$$Avg(CPP/JCS) = (1/n)\sum_{i=1}^{n}\frac{c_i}{e_i}$$

The values for the four indicators proposed above [*R*(100), *R*(6), *R*(100,k), and *R*(6,k) with their respective error terms] follow in the next columns. The consequent ranks are added in each column between brackets, with one the highest value and seven the lowest. Table 2 provides both the Pearson correlations (lower triangle) and rank-order correlations (Kendall's tau-b) between these indicators.

**Table 1**: Values and ranking (between brackets) using the various indicators.

| PI | Avg(CPP)/Avg(JCS) ≈CPP/JCSm | Avg(CPP/JCS) | R(100) | R(6) | R(100,k) | R(6,k) |
|---|---|---|---|---|---|---|
| 1 | 1.99 [1] | 2.04 (±0.51) [1] | 61.17(±7.25) [4] | 2.83(±0.38) [1] | 39.71(±4.70) [7] | 1.83(±.25) [5] |
| 2 | 1.42 [3] | 1.56 (±0.16) [3] | 69.81(±4.54) [1] | 2.68(±0.22) [3] | 72.91(±4.73) [2] | 2.79(±.23) [2] |
| 3 | 1.45 [2] | 1.60 (±0.24) [2] | 69.55(±5.60) [2] | 2.77(±0.28) [2] | 43.19(±3.48) [5] | 1.72(±.17) [6] |
| 4 | 1.17 [4] | 1.32 (±0.15) [4] | 64.34(±4.58) [3] | 2.34(±0.21) [4] | 58.12(±4.14) [3] | 2.12(±.19) [3] |
| 5 | 1.03 [5] | 1.04 (±0.15) [5] | 55.49(±4.27) [5] | 2.00(±0.15) [5] | 57.95(±4.46) [4] | 2.09(±.16) [4] |
| 6 | 0.86 [6] | 1.04 (±0.11) [6] | 49.80(±3.61) [6] | 1.88(±0.15) [6] | 91.37(±6.63) [1] | 3.44(±.27) [1] |
| 7 | 0.71 [7] | 0.87 (±0.12) [7] | 46.88(±1.87) [7] | 1.72(±0.08) [7] | 42.34(±4.75) [6] | 1.55(±.10) [7] |

Note. Numbers between parentheses ( ) are the standard errors.

---

[7] Unlike CWTS (2008 and 2010) we did not correct CPP/JCSm for self-citations in Table 1 (Opthof & Leydesdorff, 2010).

**Table 2**: Rank-order correlations (Kendall's Tau-b; upper triangle) and Pearson correlations (lower triangle) between the various indicators.

|  | Avg(CPP)/ Avg(JCS) | Avg(CPP/JCS) | R(100) | R(6) | R(100,k) | R(6,k) |
|---|---|---|---|---|---|---|
| CPP/JCSm |  | 0.98 ** | 0.62 | 1.00 ** | -0.24 | -0.05 |
| Avg(CPP/JCS) | 0.99 ** |  | 0.59 | 0.98 ** | -0.20 | 0.00 |
| R(100) | 0.68 | 0.71 |  | 0.62 | 0.14 | 0.14 |
| R(6) | 0.93 * | 0.95 ** | 0.89 ** |  | -0.24 | -0.05 |
| R(100,k) | -0.38 | -0.35 | -0.15 | -0.31 |  | 0.81 * |
| R(6,k) | -0.21 | -0.18 | -0.09 | -0.17 | 0.98** |  |

Note: **. Correlation is statistically significantly different from zero at the 0.01 level (2-tailed); *. correlation is statistically significantly different from zero at the 0.05 level (2-tailed).

As was to be expected (Waltman *et al.*, in press), the two indicators based on comparing citation scores versus journal citation scores (*Avg(CPP)/Avg(JCS)* and *Avg(CPP/JCS)*) parametrically are highly and statistically significantly correlated ($r = 0.99$, $p < 0.01$; $\tau = 0.98$, $p < 0.01$). Using the six percentile rank classes [$R(6)$], the ranking of the scientists is precisely the same as with these two average-based indicators. However, $R(100)$ deviates from this shared pattern by shifting the number 1 to fourth position behind the original numbers 2, 3, and 4. This corresponds (not incidentally) with the visual impression obtained by inspecting Figure 1. $R(100)$ can also be considered as a summary statistic of the patterns shown in Figure 1. However, $R(100)$ does not correlate significantly with the two previous indicators.

**Table 3:** Number of papers published by seven scientists categorized to six percentile rank classes.

| Percentile rank class | Weight of rank | Scientist 1 | Scientist 2 | Scientist 3 | Scientist 4 | Scientist 5 | Scientist 6 | Scientist 7 | Total |
|---|---|---|---|---|---|---|---|---|---|
| <50th (bottom-50%) | 1 | 7 | 10 | 5 | 10 | 11 | 35 | 17 | 94 |
| [50th; 75th[ | 2 | 6 | 5 | 4 | 8 | 18 | 14 | 9 | 64 |
| [75th; 90th[ | 3 | 3 | 13 | 6 | 8 | 6 | 10 | 4 | 51 |
| [90th; 95th[ | 4 | 1 | 5 | 5 | 5 | 1 | 1 | 2 | 20 |
| [95th; 99th[ | 5 | 3 | 4 | 2 | 1 | 1 | 5 | 0 | 16 |
| ≥99th (top-1%) | 6 | 3 | 0 | 0 | 0 | 0 | 0 | 0 | 3 |
| Total |  | 23 | 37 | 22 | 32 | 37 | 65 | 32 | 248 |

Note: We did not calculate Pearson's $\chi^2$-test and standardized residuals suggested by Bornmann (2010) and Bornmann and Mutz (2011) since the requirements for the calculation of the $\chi^2$-test are not fulfilled (the requirements are as follows: nearly 80% of the expected frequencies are greater than 5 and none is smaller than 1).

Table 3 informs us about the distributions of the seven document sets across the six percentile rank classes. Scientist 2 has more papers than Scientist 1 in most classes, but not in the top-1% (class 6). Given the smaller size of the *œuvre* of Scientist 1, the three papers in this class weigh heavily, namely: each for 6 * (1/23) = 0.26. This leads to a contribution of 0.78 (= 3 * 0.26) on a score of 2.83 (see Table 4, column 5, rows 7 and 8).

**Table 4**: Derivation of *R*(6) and *R*(6,k) in the case of 23 articles and proceedings papers, (co-)authored by Scientist 1 and published during 1997-2006.

| Percentile rank class ($r$) | Weight of rank ($w_r$) | $n_r$ of documents | $p_r = n / \sum_{r=1}^{6} n_r$ (normalization over the subset: $\sum_{r=1}^{6} n_r = 23$) | $(w_r * p_r)$ | $p_{r,k} = n / \sum_{r=1}^{6}\sum_{k=1}^{7} n_{r,k}$ (normalization over the set: $\sum_{r=1}^{6}\sum_{k=1}^{7} n_{r,k} = 248$) | $(w_r * p_{r,k}) * 7$ |
|---|---|---|---|---|---|---|
| <50th | 1 | 7 | 0.3043 | 0.3043 | 0.0282 | 0.1976 |
| [50th; 75th[ | 2 | 6 | 0.2609 | 0.5217 | 0.0242 | 0.3387 |
| [75th; 90th[ | 3 | 3 | 0.1304 | 0.3913 | 0.0121 | 0.2540 |
| [90th; 95th[ | 4 | 1 | 0.0435 | 0.1739 | 0.0040 | 0.1129 |
| [95th; 99th[ | 5 | 3 | 0.1304 | 0.6522 | 0.0121 | 0.4234 |
| ≥99th | 6 | 3 | 0.1304 | 0.7826 | 0.0121 | 0.5081 |
| | | 23 | $\sum_{r=1}^{6} p_r = 1$ | **R(6) = 2.8261** | $\sum_{r=1}^{6} p_{r,k=1} = (1/7)$ | **R(6,k) = 1.8347** |

Note: *k* is the number of document sets (*k* = 7).

More dramatically, however, is the difference between these two scientists when scoring in the class [95th; 99th[. The three papers of Scientist 1 in this class contribute 5 * (1/23) * 3 = 0.65 (23.0%; Table 4, row 6) to the score, while the four papers of Scientist 2 in this same class contribute only 5 * (1/37) * 4 = 0.54 (20.1%) to his/her score. The example proves our point that citation scores that do not take publication rates into account "punish" productivity because higher numbers (in the denominator) can lead to a lower appreciation of papers in similar or even higher percentile rank classes.

The use of percentile rank classes made this possible underestimation of the impact of more productive researchers quantitatively visible. However, the same effect can be expected using average-based citation scores because these indicators also operate on probability distributions while assuming independence among the samples. After such a normalization (e.g., using $z$-scores; cf. Radicchi *et al*., 2008) at the level of independent samples, the differences in size among the sets are manifested only in terms of significance testing and error terms because in these computations the $n$ of cases in each sample plays a role in the denominator (for example, as the square root of $n$ in the case of computing the standard error of the measurement). However, citation analysts hitherto have paid insufficient attention to the question whether observed differences are also statistically significant; one rarely finds error estimates in the tables or error bars in the accompanying figures and graphs.

Table 1, for example, contains the standard error of the measurement for the indicator proposed by Opthof & Leydesdorff (2010). The larger size of the error of the measurement for Scientist 1

(± 0.51) when compared with all others, and the relatively low value of this parameter for Scientist 6 (± 0.11), could have flagged this spurious publication effect of the sample sizes ($n_1$ = 23 and $n_6$ = 65, respectively). However, the last two columns of Table 1 show the effects of this correction quantitatively: Scientist 1 becomes the seventh in rank using $R(100,k)$ and fifth in rank using $R(6,k)$, whereas Scientist 6 is now the highest ranked one using either these indicators.

Using $R(6,k)$ or $R(100,k)$, the four papers of Scientist 2 in the class [95th; 99th[ (class 5) and the three papers of Scientist 1 in this same class contribute proportionally, that is, 4:3, to their respective scores. As noted in the case of $R(100,k)$ two papers in the $39^{th}$ percentile of one scientist weigh as much as one paper in the $78^{th}$ percentile of another. This is transformed in the case of using $R(6,k)$ for normative reasons; for example, because there may be more interest in the most highly-cited papers when comparing nations or institutions.

Let us repeat—in order to avoid misunderstanding—that also a weighing in terms of hundred percentile classes implies a normative decision (which may be made explicit by the specific analysis or not). The decision to weigh one paper in the $78^{th}$ percentile equally to six in the $13^{th}$ percentile (in the case of $R(100)$ and $R(100,k)$) is as normative as the choice for a scheme with six classes ($R(6)$ or $R(6,k)$) in which this ratio would only be 1:3. The weighing thus is a technical consequence of the counting on normative grounds. Our argument is that three dimensions can be analytically distinguished and therefore three separate decisions are possible in these evaluation procedures.

The transparency of *R*(100,k) can be considered as an advantage, but a six-point scale such as *R*(6,k) may be felt as more functional to communications in the policy domain. Of course, the user (e.g., the policy maker) can suggest another scheme such as *R*(5,k) by specifying other classes. Some countries use five point scales in the evaluation (e.g., the Netherlands), while in other countries six is the highest score (e.g., Germany). One may also wish to introduce a weighing scale which includes a class with weight zero for all paper below a certain threshold.

**More refined statistics**

The basic matrices which are used as inputs to the refined statistics are the one in Table 3 and a similar one for the hundred percentile classes versus seven scientists. Each paper ($n = 248$) can be attributed with a group membership, a percentile, and a classification in one of the six or hundred categories. This basic data is not affected by the normalization implied when relative frequencies are considered for each vector separately [*R*(6) and *R*(100)] or for the set as a single sample with subsets [*R*(6,k) and *R*(100,k)]. Because the citation impact classes are attributed non-parametrically, we shall use non-parametric statistics. The tests were specified above (in the methods section). Table 5 provides the results.

**Table 5**
Results for the journal-normalized citation impact comparison of seven scientists based on 100%-percentiles and six percentile rank classes

| PI | N | 100%-percentiles$ | | | | R(100) | Signed-rank Test ($H_0$: 50%) | Percentile rank classes# | | | |
|---|---|---|---|---|---|---|---|---|---|---|---|
| | | M | MDN | 95% Confidence Interval of M | | | | R(6) | $\chi^2$ test | 95% Confidence Interval of R(6) | |
| | | | | Lower | Upper | | | | | Lower | Upper |
| 1 | 23 | 60.59(±7.25) | 68.75 | 50.48 | 70.70 | 61.17(±7.25) [4] | 1.49 | 2.83(±0.38) [1] | 39.91* | 2.06 | 3.58 |
| 2 | 37 | 69.30(±4.54) | 81.07 | 62.96 | 75.63 | 69.81(±4.54) [1] | 3.40* | 2.68(±0.22) [3] | 25.51* | 2.27 | 3.13 |
| 3 | 22 | 69.00(±5.60) | 81.52 | 61.19 | 76.83 | 69.55(±5.60) [2] | 2.78* | 2.77(±0.28) [2] | 21.41* | 2.21 | 3.33 |
| 4 | 32 | 63.69(±4.58) | 63.89 | 57.29 | 70.08 | 64.34(±4.58) [3] | 2.52* | 2.34(±0.21) [4] | 11.67* | 1.91 | 2.77 |
| 5 | 37 | 54.95(±4.27) | 61.74 | 48.99 | 60.91 | 55.49(±4.27) [5] | 1.29 | 2.00(±0.15) [5] | 11.90* | 1.71 | 2.29 |
| 6 | 65 | 49.32(±3.61) | 46.72 | 44.28 | 54.37 | 49.80(±3.61) [6] | 0.03 | 1.88(±0.15) [6] | 4.93 | 1.59 | 2.17 |
| 7 | 32 | 46.39(±1.87) | 40.33 | 39.04 | 53.73 | 46.88(±1.87) [7] | -0.77 | 1.72(±0.08) [7] | 0.42 | 1.41 | 2.04 |

Notes. N=number of papers per scientist, M=mean value (± the standard error), MDN=median, ranking= ranking of the scientists' values (highest value=1). The 95%-confidence intervals of M are Goldstein-adjusted (Goldstein & Healy, 1995). The Wilcoxon signed-rank test tests the null hypothesis that the median value is 50 ("medium performance"). The $\chi^2$-test (df=5) tests the null-hypothesis that the frequency distribution over the six percentile impact classes is equal to the expected distribution (reflecting "medium performance").
* p <.05
$ Kruskal Wallis–test $\chi^2$(df=6)=20.58; * p<.05, Cramér's V=.25.[8]
# Kruskal Wallis–test $\chi^2$(df=6)=22.59; * p<.05

---

[8] A Cramér's V value of .25 (calculated on the base of the 100%-percentile rank classes) can be considered as a medium-sized effect for the association between citation performance and individual scientists. Kraemer *et al*. (2003) considered a medium-sized effect typical for the applied behavioral sciences.

The Kruskal Wallis-tests (see the legend of Table 5) calculated on the base of the 100%-percentiles and the six percentile rank classes both indicate statistically significant differences in citation impact among the seven scientists under study. The Wilcoxon signed-rank test can be used to test whether the median of each scientist's 100%-percentiles' distribution is equal to 50. This hypothesis is rejected in the case of Scientists 2, 3, and 4 ($p < 0.05$): these scientists perform above the median level (see Figure 1). Whereas Leydesdorff & Opthof (2010) found a divide between the upper four from the lower three using fractional counting, Scientist 1 can no longer be considered part of the top group after testing the 100%-percentile ranks. This accords with the above discussion about the rank-order position of Scientist 1 in Table 1.

Using $\chi^2$ Goodness-of-Fit test as a one-tailed test[9] on the values in Table 3 for each set (that is, an observed distribution) against the expected distribution of 50:25:15:5:4:1, yields statistical significance for the first five scientists, while Scientists 6 and 7 are indicated as accomplishing not differently from the expected performance level. The statistical significance of the differences between each two document sets can be tested using Mann-Whitney's U test and/or Dunn's test. We used Dunn's test for multiple comparisons with the family-wise adjusted alpha as the significance level (0.05/21 = 0.0023; Table 6).

---

[9] These tests can be performed, for example, at http://faculty.vassar.edu/lowry/csfit.html.

**Table 6**: Dunn's post-hoc test (family-adjusted α = 0.05/21 = 0.0023) for hundred percentile classes in the upper triangle and six percentile ranks in the lower triangle. Confidence levels (at the 99.77%-level) are only shown for the significant comparisons (*p* < 0.0023).

| | Scientist 1 | 2 | 3 | 4 | 5 | 6 | 7 |
|---|---|---|---|---|---|---|---|
| 1 | | | | | | | |
| 2 | | | | | | 1.91 - 38.04 | 1.73 - 44.08 |
| 3 | | | | | | | |
| 4 | | | | | | | |
| 5 | | | | | | | |
| 6 | 0.03 - 1.87 | 0.02 - 1.58 | | | | | |
| 7 | 0.07 -2.14 | 0.04 - 1.87 | 0.01-2.10 | | | | |

The statistical significance of the differences between Scientists 2, on the one side, and 6 and 7, on the other, is indicated in both triangles. The difference between Scientists 3 and 7 is indicated as significant when using the six percentile rank classes, and the significance of the differences between Scientist 1 versus both numbers 6 and 7 are also flagged in this case. The aggregation thus can enhance the visibility of differences among groups.[10]

In summary, several tests are available to study the statistical significance of such differences in greater detail. The performances of these seven scientists are different (Kruskall-Wallis), and some differences are indicated as more robust (in terms of significance and/or confidence intervals) against changing the percentile rank classes from hundred to six. Aggregation into six percentile rank classes makes differences among these sets more pronounced. In this section, we mainly wished to show that a fine-grained statistical apparatus is available to study the statistical significance of differences in these rankings in detail. Only with statistical methods is it possible to assess performance differences between scientists, research groups etc. as meaningful or not. In our opinion, the use of rules of thumb is not sufficiently transparent and should therefore be avoided.

---

[10] Using the Tukey test (with Bonferroni *ex post* correction), one would additionally be able to test for statistically homogeneous subgroups.

Let us note that differences in the distributions are tested for significance in terms of the *data*. After the external normalization against the reference set, each score can be attributed unambiguously to a percentile rank class in the schemes of both six and hundred categories. One can also wish to test the differences in the scores (that is, $x * p_x$ as contributions to $R(x)$; see Eq. 1 above) for each set using the same tests but based on the matrix of seven sets versus 6 or 100 variables, respectively. One can test the differences in the indicator values for their statistical significance by multiplying the cell values in the matrix with $k * \sum_i n_i / \sum_k \sum_i n_{ik}$. In this study, $k = 7$ and $i = 6$ or $100$ for $R_{6,k}$ and $R_{100,k}$, respectively.

In this design, the Kruskal-Wallis rank variance test rejects—as before—the hypothesis that the distributions are the same in the case of both $R(100,k)$ and $R(6,k)$. Table 7 provides the confidence levels for these two statistics and Table 8 the results of Dunn's test for the comparisons among the seven (sub)samples.

**Table 7:** Confidence intervals for $R(100,k)$ and $R(6,k)$.

| PI | M ± SE | R(100,k) | 95% Confidence Interval Lower | 95% Confidence Interval Upper | R(6,k) | 95% Confidence Interval Lower | 95% Confidence Interval Upper |
|----|--------|----------|-------|-------|--------|-------|-------|
| 1 | 39.33(±4.70) | 39.71(±4.70) [7] | 29.87 | 48.10 | 1.83(±0.25) [5] | 1.41 | 2.38 |
| 2 | 72.37(±4.74) | 72.91(±4.73) [2] | 62.64 | 81.40 | 2.79(±0.23) [2] | 2.39 | 3.26 |
| 3 | 42,85(±3.48) | 43.19(±3.48) [5] | 36.06 | 49.33 | 1.72(±0.17) [6] | 1.38 | 2.08 |
| 4 | 57.52(±4.14) | 58.12(±4.14) [3] | 49.54 | 65.23 | 2.12(±0.19) [3] | 1.75 | 2.53 |
| 5 | 57.39(±4.46) | 57.95(±4.46) [4] | 48.23 | 65.77 | 2.09(±0.16) [4] | 1.79 | 2.42 |
| 6 | 90.49(±6.63) | 91.37(±6.63) [1] | 77.77 | 103.81 | 3.44(±0.27) [1] | 2.94 | 4.00 |
| 7 | 41.90(±4.75) | 42.34(±4.75) [6] | 32.82 | 51.47 | 1.55(±0.10) [7] | 1.28 | 1.87 |

Except for the differences caused by the rounding to integers, $R(100,k)$ is equal to the mean of the percentiles for analytical reasons. The InCite™ database of Thomson Reuters uses these means of the percentiles, but we used in this study hundred and six percentile rank classes in order to show how a normative evaluation scheme can be used in addition to the citation impact scores. However, the differences between these means and $R(100,k)$ are marginal.

**Table 8**: Dunn's post-hoc test (family-adjusted α = 0.05/21 = 0.0023) for $R_{100,k}$ in the upper triangle and $R_{6,k}$ in the lower triangle; confidence levels (at the 99.77%-level) are indicated when the multiple comparison is significant ($p < 0.0023$).

| PI | 1 | 2 | 3 | 4 | 5 | 6 | 7 |
|----|---|---|---|---|---|---|---|
| 1 |   | 4.39 – 61.68 |   |   |   | 24.98 – 77.33 |   |
| 2 |   |   | 4.76 – 58.56 |   |   |   | 4.43 – 56.51 |
| 3 |   |   |   |   |   | 21.03 – 74.25 |   |
| 4 |   |   |   |   |   | 9.67 – 56.26 |   |
| 5 |   |   |   |   |   | 10.88 – 55.32 |   |
| 6 | 0.52 – 2.70 |   | 0.61 – 2.83 | 0.35 - 2.29 | 0.43 – 2.28 |   | 25.29 – 71.89 |
| 7 |   | 0.15 – 2.33 |   |   |   | 0.92 – 2.86 |   |

On the basis of both $R(6,k)$ and $R(100,k)$, the scores for Scientist 6 are significantly different from all other scientists except Scientist 2 ($p < 0.0023$). The latter differs from Scientist 7. In general, Table 8 is differently, but much denser populated than Table 6. Table 6 was based on considering the samples as independent, while in this case the design is one of related samples.

**Conclusions and discussion**

Our purpose in this study was to develop citation impact indicators based not on averages but on percentile ranks. We specified a number of criteria for a more abstract scheme that can also be used to organize and/or schematize different citation impact indicators according to three degrees of freedom: the selection of the reference sets, the evaluation criteria, and the choice of whether or not to define the samples as independent.

The proposed indicators [$R(6)$, $R(100)$, $R(6,k)$, $R(100,k)$] are an improvement of averages-based indicators first because using non-parametric statistics one can abstract from the shape of the distribution of citations over papers. Secondly, the choice of the reference set for each paper is no longer related to the evaluation scheme. Both the reference sets can be chosen—for example, as individual journals, groups of journals (e.g., ISI Subject Categories), papers selected on specific criteria such as index terms or keywords, etc.—and evaluation schemes specified, in terms of six classes or otherwise. The latter choice is a normative one, while the choice of external reference sets is in need of analytical grounding using, for example, bibliometric arguments.

The elaboration of the proposal of Bornmann & Mutz (2011) to use percentile ranks made us aware how sensitive citation-based indicators can be to sample sizes even after correction for differences in the shapes of the distributions. This result provided us with the major learning step of the study: one should compare "like with like" as Martin & Irvine (1983) once formulated in the early days of citation analysis, but one should not reduce this comparison to the specification of reference set(s) for each article. The document subsets under study are to be compared among themselves after being normalized at the individual paper level against the reference sets. The normalization in terms of the external reference sets and thereafter the rewrite as percentiles was not yet sufficient, since additional normalization is needed as relative frequencies across the sets under study. By normalizing the relative frequencies in terms of the grand total of the combined sample, one eventually obtains percentile rank scores that account for differences both in the size and the shape of the citation distributions. These scores [$R(100,k)$ and $R(6,k)$]—or more generally: $R(i,k)$—are comparable across sets.

Our data provided us with an opportunity to make a convincing case for this change in the framework of citation analysis—from considering sets as independent samples to subsamples of a single sample—by showing the mistake in the evaluation that one can make when one uses citation rates without taking sample sizes into account. Only because of the smaller sample size was Scientist 1 at the top of the ranking: a paper of equal rank contributed in his/her case $1/23 = 0.043$ to the total score while it would contribute only $1/37 = 0.027$ for Scientist 2. Scientist 6 with 65 papers was disadvantaged to the extent that s/he led the ranking after we corrected for this size effect. Without this correction, the percentile ranks $R(6)$ and $R(100)$ correlated highly and significantly with the CWTS indicators (with or without the correction for the order of operations); scores based on averages suffer from this ignored size-effect.

In other words, the initial step of Lundberg (2007) and Opthof & Leydesdorff (2010) to introduce significance testing and error indication into the measurement of average citations (as had been done previously by other centers; cf. Gingras & Archambault (2011)) was not a sufficient step. The percentile rank approach of Bornmann & Mutz (2011) made it clear that the assumption hitherto of comparing independent probability distributions when using the mean or the median (or any other statistics) requires further reflection. In citation analysis, one compares samples which are not necessarily independent. Without normalization across the samples, one changes the basis for the comparison when moving from one set to another.

In summary, using these indicators the citation analyst has three degrees of freedom: (1) freedom to choose a normative evaluation scheme, (2) freedom—or in other terms, the need—to rationalize the choice for external reference sets, and (3) freedom to decide whether sample definitions are considered independent or not. These three dimensions are analytically different, and the eventual scores will differ with these choices. *Vice versa*, there are no absolute citation impact scores or world averages independent of making these choices in the design of a study. As we have argued, zero-normalizations can be used such as choosing *not* to normalize against reference sets (that is, assuming the reference values to be equal to unity). The citation impact enterprise is thoroughly probabilistic and precisely the probability distributions lead us to a strong preference for defining probabilities across sets such as when using $R(i,k)$ where $i$ is the indicator for the (percentile rank) class and $k$ the indicator of each subset. This new measure enables us to compare sets of different sizes among one another.

**References**

Albarrán, P., & Ruiz-Castillo, J. (2011). References made and citations received by scientific articles. *Journal of the American Society for Information Science and Technology, 62*(1), 40-49.

Bornmann, L. (2010). Towards an ideal method of measuring research performance: some comments to the Opthof and Leydesdorff (2010) paper. *Journal of Inormetrics, 4*(3), 441-443.

Bornmann, L. (2011). Scientific peer review. *Annual Review of Information Science and Technology, 45*, 199-245.

Bornmann, L., de Moya-Anegón, F., & Leydesdorff, L. (2010a). Do scientific advancements lean on the shoulders of giants? A bibliometric investigation of the Ortega hypothesis. *PLoS ONE, 5*(10), e11344.

Bornmann, L., Leydesdorff, L., & Van den Besselaar, P. (2010b). A Meta-evaluation of Scientific Research Proposals: Different Ways of Comparing Rejected to Awarded Applications. *Journal of Informetrics, 4*(3), 211-220.

Bornmann, L., & Mutz, R. (2011). Further steps towards an ideal method of measuring citation performance: The avoidance of citation (ratio) averages in field-normalization. *Journal of Informetrics, 5*(1), 228-230.

Bornmann, L., Mutz, R., Marx, W., Schier, H., & Daniel, H.-D. (2011). A multilevel modelling approach to investigating the predictive validity of editorial decisions: do the editors of a high-profile journal select manuscripts that are highly cited after publication? *Journal of the Royal Statistical Society - Series A (Statistics in Society), 174*(4), 1-23.

Bornmann, L., Mutz, R., Neuhaus, C., & Daniel, H.-D. (2008). Use of citation counts for research evaluation: standards of good practice for analyzing bibliometric data and presenting and interpreting results. *Ethics in Science and Environmental Politics, 8*,

93-102. doi: 10.3354/esep00084.

Bornmann, L., Schier, H., Marx, W., & Daniel, H.-D. (2011). Is Interactive Open Access Publishing Able to Identify High-Impact Submissions? A Study on the Predictive Validity of Atmospheric Chemistry and Physicsby Using Percentile Rank Classes, *Journal of the American Society for Information Science and Technology, 62*(1), 61-71.

Boyack, K. W. (2004). Mapping knowledge domains: characterizing PNAS. *Proceedings of the National Academy of Sciences of the United States of America, 101*, 5192-5199.

Boyack, K. W., Klavans, R., & Börner, K. (2005). Mapping the Backbone of Science. *Scientometrics, 64*(3), 351-374.

Cohen, J. (1988). *Statistical power analysis for the behavioral sciences* (2nd ed.). Hillsdale, NJ, USA: Lawrence Erlbaum Associates, Publishers.

Colliander, C., & Ahlgren, P. (2011). The effects and their stability of field normalization baseline on relative performance with respect to citation impact: a case study of 20 natural science departments. *Journal of Informetrics, 5*(1), 101-113.

Corder, G. W., & Foreman, D. I. (2009). *Nonparametric statistics for non-statisticians*. New York, NY, USA: Wiley.

Cramér, H. (1980). *Mathematical methods of statistics*. Princeton, NJ, USA: Princeton University Press.

CWTS. (2008). AMC-specifieke CWTS-analyse 1997–2006 (access via AMC intranet; unpublished, confidential). Leiden, The Netherlands: CWTS.

CWTS. (2010). AMC-specifieke CWTS-analyse 1997–2008 (access via AMC intranet; unpublished, confidential). Leiden, The Netherlands: CWTS.

Evidence Ltd. (2007). The use of bibliometrics to measure research quality in UK higher education institutions. London, UK: Universities UK.

Garfield, E. (1971). The mystery of the transposed journal lists—wherein Bradford's Law of Scattering is generalized according to Garfield's Law of Concentration. *Current Contents, 3*(33), 5–6.

Garfield, E. (1979). Is citation analysis a legitimate evaluation tool? *Scientometrics, 1*(4), 359-375.

Gingras, Y., & Larivière, V. (2011). There are neither "king" nor "crown" in scientometrics: Comments on a supposed "alternative" method of normalization. *Journal of Informetrics, 5*(1), 226-227.

Glänzel, W. (1992). Publication Dynamics and Citation Impact: A Multi-Dimensional Approach to Scientometric Research Evaluation. In P. Weingart, R. Sehringer & M. Winterhagen (Eds.), *Representations of Science and Technology. Proceedings of the International Conference on Science and Technology Indicators, Bielefeld, 10-12 June 1990* (pp. 209-224). Leiden: DSWO / Leiden University Press.

Glänzel, W. (2010). On reliability and robustness of scientometrics indicators based on stochastic models. An evidence-based opinion paper. *Journal of Informetrics, 4*(3), 313-319.

Glänzel, W., & Schubert, A. (2003). A new classification scheme of science fields and subfields designed for scientometric evaluation purposes. *Scientometrics, 56*(3), 357-367.

Glänzel, W., Thijs, B., Schubert, A., & Debackere, K. (2009). Subfield-specific normalized relative indicators and a new generation of relational charts: Methodological foundations illustrated on the assessment of institutional research performance.

*Scientometrics, 78*(1), 165-188.
Goldstein, H., & Healy, M. J. R. (1995). The graphical presentation of a collection of means. *Journal of the Royal Statistical Society. Series A (Statistics in Society), 158*(1), 175-177.
Hirsch, J. E. (2005). An index to quantify an individual's scientific research output. *Proceedings of the National Academy of Sciences of the United States of America, 102*(46), 16569-16572.
Jackson, C. (1996). *Understanding psychological testing*. Leicester, UK: British Psychological Society.
Joint Committee on Quantitative Assessment of Research. (2008). Citation statistics. A report from the International Mathematical Union (IMU) in cooperation with the International Council of Industrial and Applied Mathematics (ICIAM) and the Institute of Mathematical Statistics (IMS). Berlin, Germany: International Mathematical Union (IMU).
Kline, R. B. (2004). *Beyond significance testing: reforming data analysis methods in behavioral research*. Washington, DC, USA: American Psychological Association.
Kraemer, H. C., Morgan, G. A., Leech, N. L., Gliner, J. A., Vaske, J. J., & Harmon, R. J. (2003). Measures of clinical significance. *Journal of the American Academy of Child and Adolescent Psychiatry, 42*(12), 1524-1529. doi: DOI 10.1097/01.chi.0000091507.46853.d1.
Kurtz, M. J., & Henneken, E. A. (2007). Open Access does not increase citations for research articles from *The Astrophysical Journal*. Retrieved April 11, 2011, from http://arxiv.org/abs/0709.0896
Kvam, P. H., & Vidakovic, B. (2007). *Nonparametric statistics with applications to science and engineering*. Hoboken, NJ, USA: Wiley-Interscience.
Levine, G. (1991). *A Guide to SPSS for Analysis of Variance*. Hillsdale, NJ: Lawrence Erlbaum.
Leydesdorff, L. (2008). *Caveats* for the Use of Citation Indicators in Research and Journal Evaluation. *Journal of the American Society for Information Science and Technology, 59*(2), 278-287.
Leydesdorff, L., & Bornmann, L. (2011). How fractional counting affects the Impact Factor: Normalization in terms of differences in citation potentials among fields of science. *Journal of the American Society for Information Science and Technology 62*(2) 217-229
Leydesdorff, L., & Opthof, T. (2010). Normalization at the field level: fractional counting of citations. *Journal of Informetrics, 4*(4), 644-646.
Leydesdorff, L., & Opthof, T. (2011). Remaining problems with the "New Crown Indicator" (MNCS) of the CWTS. *Journal of Informetrics, 5*(1), 224-225.
Leydesdorff, L., & Rafols, I. (2009). A Global Map of Science Based on the ISI Subject Categories. *Journal of the American Society for Information Science and Technology, 60*(2), 348-362.
Lundberg, J. (2007). Lifting the crown - citation *z*-score. *Journal of Informetrics, 1*(2), 145-154.
Martin, B., & Irvine, J. (1983). Assessing Basic Research: Some Partial Indicators of Scientific Progress in Radio Astronomy. *Research Policy, 12*, 61-90.
Moed, H. F. (2010). Measuring contextual citation impact of scientific journals. *Journal of Informetrics, 4*(3), 265-277.

National Science Board (2010). Science and engineering indicators 2010, appendix tables. Arlington, VA, USA: National Science Foundation (National Science Board 10-01).

Opthof, T., & Leydesdorff, L. (2010). Caveats for the journal and field normalizations in the CWTS ("Leiden") evaluations of research performance. *Journal of Informetrics, 4*(3), 423-430.

Opthof, T., & Leydesdorff, L. (in preparation). Citation analysis cannot legitimate the strategic selection of excellence.

Pudovkin, A. I., & Garfield, E. (2002). Algorithmic procedure for finding semantically related journals. *Journal of the American Society for Information Science and Technology, 53*(13), 1113-1119.

Pudovkin, A. I., & Garfield, E. (2009). Percentile Rank and Author Superiority Indexes for Evaluating Individual Journal Articles and the Author's Overall Citation Performance. *CollNet Journal of Scientometrics and Information Management, 3*(2), 3-10.

Radicchi, F., Fortunato, S., & Castellano, C. (2008). Universality of citation distributions: Toward an objective measure of scientific impact. *Proceedings of the National Academy of Sciences, 105*(45), 17268-17272.

Rafols, I., & Leydesdorff, L. (2009). Content-based and Algorithmic Classifications of Journals: Perspectives on the Dynamics of Scientific Communication and Indexer Effects *Journal of the American Society for Information Science and Technology, 60*(9), 1823-1835.

Rons, N., & Amez, L. (2008). Impact Vitality - a measure for excellent scientists. In J. Gorraiz & E. Schiebel (Eds.), *Excellence and emergence. A new challenge for the combination of quantitative and qualitative approaches. 10th International Conference on Science and Technology Indicators* (pp. 211-213). Vienna, Austria: Austrian Research Centers (ARC).

Ross, S. M. (2007). *Introduction to probability models*. London, UK: Elsevier.

Rousseau, R., & Leydesdorff, L. (2011). Simple arithmetic versus intuitive understanding: The case of the impact factor. *ISSI Newsletter* (in press).

Sheskin, D. (2007). *Handbook of parametric and nonparametric statistical procedures* (4th ed.). Boca Raton, FL, USA: Chapman & Hall/CRC.

Schubert, A., & Braun, T. (1993). Reference standards for citation based assessments. *Scientometrics, 26*(1), 21-35.

Schubert, A., & Braun, T. (1996). Cross-field normalization of scientometric indicators. *Scientometrics, 36*(3), 311-324.

Schubert, A., & Glänzel, W. (1983). Statistical reliability of comparisons based on the citation impact of scientific publications. *Scientometrics, 5*(1), 59-73.

Seglen, P. O. (1992). The Skewness of Science. *Journal of the American Society for Information Science, 43*(9), 628-638.

Sheskin, D. (2007). *Handbook of parametric and nonparametric statistical procedures* (4th ed.). Boca Raton, FL, USA: Chapman & Hall/CRC.

Spaan, J. A. E. (2010). The danger of pseudoscience in informetrics. *Journal of Informetrics 4*(3), 439-440.

Van Raan, A. F. J. (2005). Measurement of central aspects of scientific research: performance, interdisciplinarity, structure. *Measurement, 3*(1), 1-19.

Van Raan, T. (2010). Bibliometrics: measure for measure. [10.1038/468763a]. *Nature, 468*(7325), 763-763.

Van Raan, A. F. J., Van Leeuwen, T. N., Visser, M. S., Van Eck, N. J., & Waltman, L.

(2010a). Rivals for the crown: reply to Opthof and Leydesdorff. *Journal of Informetrics, 4*(3), 431-435.

Van Raan, A. F. J., Eck, N. J. v., Leeuwen, T. N. v., Visser, M. S., & Waltman, L. (2010b). *The new set of bibliometric indicators of CWTS.* Paper presented at the 11th International Conference on Science and Technology Indicators, Leiden, September 9-11, 2010; pp. 291-293.

Waltman, L., Van Eck, N. J., Van Leeuwen, T. N., Visser, M. S., & Van Raan, A. F. J. (2011). Towards a New Crown Indicator: Some Theoretical Considerations. *Journal of Informetrics, 5*(1), 37-47.

Waltman, L., van Eck, N. J., van Leeuwen, T. N., Visser, M. S., & van Raan, A. F. J. (in press). Towards a new crown indicator: An empirical analysis. *Scientometrics.* Arxiv preprint arXiv:1004.1632.

Zitt, M. (2010). Citing-side normalization of journal impact: A robust variant of the Audience Factor. *Journal of Informetrics, 4*(3), 392-406.

Zitt, M., & Small, H. (2008). Modifying the journal impact factor by fractional citation weighting: The audience factor. *Journal of the American Society for Information Science and Technology, 59*(11), 1856-1860.